# A Stackelberg Strategy for Routing Flow over Time

Umang Bhaskar[*]    Lisa Fleischer[*]    Elliot Anshelevich[†]

**Abstract**

Routing games are used to to understand the impact of individual users' decisions on network efficiency. Most prior work on routing games uses a simplified model of network flow where all flow exists simultaneously, and users care about either their maximum delay or their total delay. Both of these measures are surrogates for measuring how long it takes to get all of a user's traffic through the network. We attempt a more direct study of how competition affects network efficiency by examining routing games in a flow over time model. We give an efficiently computable Stackelberg strategy for this model and show that the competitive equilibrium under this strategy is no worse than a small constant times the optimal, for two natural measures of optimality.

[*]{umang,lkf}@cs.dartmouth.edu. Partially supported by NSF grants CCF-0728869 and CCF-1016778.
[†]eanshel@cs.rpi.edu. Partially supported by NSF grants CCF-0914782 and NetSE-1017932.

# 1 Introduction

In routing games, players route a fixed amount of flow in a network. A player suffers a cost, which depends on its routing and the routing chosen by the other players. A flow in a routing game is an equilibrium flow if no player can choose a different routing and reduce its cost.

Routing games model a variety of problems, including routing on roads [3, 32], computer networks [10, 24, 28], and scheduling tasks on machines [22]. For many measures of the quality of a routing, the equilibrium in routing games is known to be inefficient compared to a routing which optimizes the measure. This inefficiency is quantified by the *price of anarchy* [25]: the worst ratio of the objective evaluated for the equilibrium flow, to the optimal flow. There is considerable interest in obtaining bounds on the price of anarchy in routing games.

Players in a routing game have a *bottleneck* objective if a player's cost is the maximum delay on the edges it uses [5]. The bottleneck objective models applications where a player's cost depends largely on the performance of the worst resource it uses. This objective ignores the effect of delay on edges besides the bottleneck edge, which can lead to the counterintuitive situation where players fail to distinguish between two strategies which have the same bottleneck, but have considerably different delays. This behavior may result in an unbounded price of anarchy, e.g., [2, 5, 10]. In many of these bad instances, the price of anarchy would be 1 if player costs depended on edges besides the bottleneck edge. Models where a player takes into account the delays on all edges have an improved price of anarchy [10].

Even models where a player's cost is an aggregation of the cost on each edge assume that the flow is static: every edge has flow on it instantaneously and simultaneously and, once established, a flow continues indefinitely. However, the flow in networks is often transient. Flow enters a network, uses it for some time, and then exits, and the flow on an edge changes with time.

This time-varying nature of flows is captured by flows over time, introduced in [14]. In this model, flow traverses the path in finite time, and exits the network at the sink. Thus, the flow on each edge of the network varies with time. Every edge has a capacity which limits the flow rate on the edge.

We consider routing games for flows over time. Every player controls infinitesimal flow, and in contrast to previous models where users care about either their maximum delay or their total delay, in our model the cost of a player is the time at which it arrives at the sink. A player's strategy is a path from the source to the destination. On every edge, the flow follows first-in, first-out (FIFO). While the network is capacitated, the model allows the inflow on an edge to be larger than the capacity of the edge[‡]. The excess flow forms a queue at the tail of the edge, and must wait for the preceding flow to exit before it can exit the edge. Although the capacities and the edge-delays are fixed, the total delay on an edge varies with the size of the queue on the edge. The queue size seen by flow arriving at the edge varies with time; hence the total delay along any path varies with time.

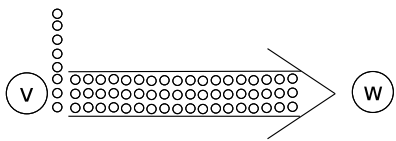

Figure 1: Flow in excess of capacity on an edge forms a queue at the tail

This game, which we call a *temporal routing game*, follows the model of selfish routing of flows over time used in [20]. The model possesses a number of interesting characteristics. It follows FIFO, which is a standard assumption in traffic routing literature. The model is based on dynamic queueing, first used in [33].

---

[‡]Thus, unlike the static flow model considered in [11], the ability to reduce capacities is not sufficient to enforce the optimal flow.

Further, the equilibrium flow over time can be characterized in terms of special static flows [20], described in §2.

In flows over time, similar to static flows, various objectives may be used to compare the performance of the equilibrium flow to the optimal flow. A natural objective is the total delay: the sum of the costs of the players. The flow which minimizes the total delay is the *earliest arrival flow*, which maximizes the flow that arrives at the destination by time $\theta$, for every time $\theta$. We call the ratio of the total delay of the worst equilibrium flow to the total delay of earliest arrival flow the *total delay price of anarchy*. A different objective is the time taken to route a fixed amount of flow to the sink. This is called the completion time, and is minimized by a *quickest flow*. The earliest arrival flow is also a quickest flow. The ratio of the time taken by the worst equilibrium flow, to the time taken by the quickest flow to route a fixed amount of flow is called the *time price of anarchy*. A third objective is the amount of flow that reaches the destination by time $\theta$. The ratio of amount of flow which reaches the destination by time $\theta$ in the worst equilibrium flow to the amount of flow which reaches the sink in the earliest arrival flow is called the *evacuation price of anarchy*.

In the *Stackelberg model* introduced in [31], different players in a game have different priorities. A *leader* picks a strategy first, and then the followers pick their strategies. Importantly, the leader commits to a strategy before the followers pick theirs. In their 1982 book on noncooperative game theory [6], Basar and Olsder describe a general form of Stacklberg games where players may have different strategy spaces. We embrace this general definition. In our setting, the network manager is the leader. Given some physical limit on the capacity of each edge, the network manager acting as leader picks a capacity for each edge which does not exceed this physical limit. The remaining players, acting as followers, then pick a route from source to sink as their strategy.

**Our Contribution.** We study the equilibrium flow in temporal routing games. We show that small constant bounds on the efficiency loss of equilibrium flow in temporal routing games can be enforced. In particular,

- We give a polynomial-time computable Stackelberg strategy to enforce a bound of $e/(e-1)$ on the time price of anarchy in temporal routing games; and

- We show the same strategy also enforces a bound of $2e/(e-1)$ on the total delay price of anarchy.

The strategy we describe is based on the following key result.

- In temporal routing games where the edge capacities satisfy certain properties with respect to the quickest flow, the time price of anarchy is bounded by $e/(e-1)$.

The bound of $e/(e-1)$ on the time price of anarchy stated above is tight, as there is a matching example in [20]. Our results are in sharp contrast to two previous results. In [20], the authors show that the evacuation price of anarchy is $\Theta(\log n)$, where $n$ is the number of vertices. Further, [23] considers the maximum time taken by any player to travel through the network. They show that for this objective, the price of anarchy is $\Omega(n)$.

We restrict our analysis to single-source, single-sink networks with constant inflow. In this case, the quickest flow is known to be a temporally repeated flow (see §2). There are variable inflows for which this is not true. Equilibrium for temporal routing games with a single source and single sink exist [20]. Further, in settings with constant inflow, they can be described in terms of static flows with special properties [20]. These properties are crucially used in our proofs.

**Related Work.** For selfish routing of static flows, the literature is vast; see [7, 26, 32] for early results on the equilibrium in selfish routing. The term price of anarchy was first used in [25] to describe the efficiency loss caused by the absence of a controlling authority. Since then, the price of anarchy has been widely used as a measure of how system performance degrades if resources are used selfishly. For results on the price of anarchy for static flows, see [28].

Stackelberg strategies have been used in computer science literature to manage the efficiency loss at equilibrium [21, 27, 29]. Here, the network manager is a player with flow which she routes with the objective of reducing efficiency loss. *Coordination mechanisms*, introduced in [8], refer to a choice of system parameters by the designer to influence equilibria. The term is used to describe situations both where the system parameters are chosen before the market power of the players is known, and where the market power of the players is known beforehand. In the latter situation, this concept fits within the notion of a Stackelberg game as defined by Basar and Olsder [6]. Our approach in this paper may also be viewed as a coordination mechanism with the market power of players known. Coordination mechanisms and related results are further discussed in [9].

Ford and Fulkerson [14] introduce flows over time. They consider the problem of maximizing the amount of the flow which can be sent from a source $s$ to a destination $t$ by a given time $T$; the flow which achieves this is called the maximum dynamic flow. A related problem is the quickest flow problem: find the dynamic flow which routes a fixed amount of flow $M$ from $s$ to $t$ in minimum time. The earliest arrival flow problem generalizes the maximum flow and the quickest flow problems. For a single source and destination, earliest arrival flows exist [15], however the flow over time obtained may have a description of size exponential in the size of the input [34]. These and other problems on flows over time are considered in [12, 13, 17, 18, 19].

Temporal routing games are analyzed by Koch and Skutella in [20] and are based on deterministic queueing models used earlier in traffic simulation [30, 33]. The authors in [20] show that if all edges have zero delay, the time price of anarchy is 1. In contrast, they show the evacuation price of anarchy is $\Theta(\log n)$. Macko et al. study the existence of Braess's paradox in temporal routing games [23]. They show that the maximum delay suffered by any player can be arbitrarily worse than for an optimal flow over time which minimizes the maximum delay. Anshelevich and Ukkusuri [4] analyze a different discrete-time model of selfish routing. In their model, the delay of an edge $e$ at any timestep $t$ is a function of the flow entering the edge at $t$ and the *history* of the edge, which is an encoding of the flow entering the edge in timesteps before $t$. However, in their model, edges are uncapacitated; and in instances with multiple sources and sinks, the equilibrium may not exist. In single-source, single-sink instances an equilibrium exists and can be computed efficiently. However the time price of anarchy may be large. Hoefer et al. [16] consider a different model where they consider flow controlled by players as tasks and edges as machines. A player's task corresponds to a significant amount of flow, rather than infinitesimal as in the models discussed previously, and must be routed on a single path.

## 2  Model, Notation and Definitions

Let $G = (V, E)$ be a directed acyclic graph with two special vertices $s$ and $t$ called the source and sink. Each edge in the graph has a nonnegative capacity $c_e$ and a nonnegative edge-delay $d_e$. An $s$-$t$ path in the graph is a sequence of edges $(v_0, w_0), \ldots, (v_l, w_l)$ such that $v_0 = s$, $w_l = t$, $w_i = v_{i+1}$ and $v_i \neq v_j$ for $i \neq j$. We abuse notation slightly and define $d_p := \sum_{e \in p} d_e$ for any path $p$.

**Static Flows.** In a directed graph $G = (V, E)$ with capacities $c_e$ on the edges and source $s$ and sink $t$, a static flow $f$ is an assigment of nonnegative values $f_{uv}$ that satisfy $f_{uv} = 0$ for $(u, v) \notin E$ and capacity constraints (1) and flow conservation (2) for $(u, v) \in E$:

$$f_e \leq c_e \qquad \forall e \in E \qquad (1)$$

$$\sum_u f_{uv} = \sum_w f_{vw} \qquad \forall v \in V \setminus \{s, t\} \qquad (2)$$

The *value* of a static flow $f$ is $|f| = \sum_v f_{sv}$. A *path flow* $f_p$ on a path $p$ is a flow on $p$ of value $|f_p|$. For an acyclic graph, a static flow $f$ can be decomposed into the sum of path flows on a set of paths $\mathcal{P}$ so that $f_e = \sum_{p \in \mathcal{P}: e \in p} |f_p|$ [1]. We use $f = \{f_p\}_{p \in \mathcal{P}}$ to denote a flow decomposition of flow $f$, where $\mathcal{P}$ is the set of paths with strictly positive flow.

**Flows over Time.** A flow over time is denoted $(f^+, f^-)$ and is defined by the functions of time $f_{uv}^+$ and $f_{uv}^-$, $\forall u, v \in V$. For any time $\theta \in \mathbb{R}_+$ and $(u, v) \notin E$, $f_{uv}^+(\theta) = f_{uv}^-(\theta) = 0$. For $e = (u, v) \in E$ and time $\theta \in \mathbb{R}_+$, $f_e^+(\theta)$ is the *rate* of flow into edge $e$ at time $\theta$, and $f_e^-(\theta)$ is the rate of flow out of edge $e$ at time $\theta$. A flow over time $(f^+, f^-)$ is feasible if it satisfies capacity constraints (3) and flow conservation (4):

$$f_e^-(\theta) \leq c_e \qquad \forall e \in E,\, \theta \in \mathbb{R}_+ \qquad (3)$$

$$\sum_u f_{uv}^-(\theta) = \sum_w f_{vw}^+(\theta) \qquad \forall v \in V \setminus \{s, t\}, \qquad (4)$$

$$\forall \theta \in \mathbb{R}_+$$

The net flow rate leaving $s$ and entering $t$ must be positive.

$$\sum_u f_{us}^+(\theta) - \sum_w f_{sw}^-(\theta) \leq 0 \quad \forall \theta \in \mathbb{R}_+ \qquad (5)$$

$$\sum_u f_{ut}^+(\theta) - \sum_w f_{tw}^-(\theta) \geq 0 \quad \forall \theta \in \mathbb{R}_+ \qquad (6)$$

For a vertex $v$, define $f_v^+(\theta) := \sum_u f_{uv}^-(\theta)$ and $f_v^-(\theta) := \sum_w f_{vw}^+(\theta)$.

The rate of flow entering an edge $f_e^+(\theta)$ may be larger than the capacity of the edge. In this case, the excess flow forms a *queue* at the tail of the edge and must wait for the flow before it in the queue, before it starts traversing the edge. Define the total flow entering and exiting edge $e$ by time $\theta$ to be $F_e^+(\theta) = \int_0^\theta f_e^+(\nu) d\nu$ and $F_e^-(\theta) = \int_0^\theta f_e^-(\nu) d\nu$ respectively. The edge-delay $d_e$ is the time taken by flow to traverse the edge if there is no queue on the edge. Then $\forall e \in E,\, \theta \in \mathbb{R}_+$,

$$F_e^-(\theta) \leq F_e^+(\theta - d_e). \qquad (7)$$

To ensure that flow entering an edge at any time also leaves the edge after finite time, the flow over time must satisy, $\forall e \in E, \theta \in \mathbb{R}_+$,

$$\exists \triangle < \infty : \; F_e^+(\theta) \leq F_e^-(\theta + d_e + \triangle). \qquad (8)$$

The *queuing-delay* $q_e(\theta)$ on edge $e$ at time $\theta$ is the minimum time flow entering the edge at time $\theta$ must wait before it starts traversing the edge:

$$q_e(\theta) := \min\{\triangle \geq 0 : F_e^+(\theta) = F_e^-(\theta + d_e + \triangle)\} \qquad (9)$$

When flow leaves the queue, the time taken to traverse the edge is the edge-delay $d_e$. The *total-delay* at $e$ of flow entering edge $e$ at time $\theta$ is $d_e + q_e(\theta)$. By (9), flow entering an edge at time $\theta$ must allow all the flow which entered earlier to exit the edge before it can exit, hence flow on an edge follows FIFO.

Flow enters the graph at the source $s$ at a constant rate $c_0$. The total amount of flow to be routed to the sink is $M$. The *completion time* is the time when $M$ units of flow arrive at the sink.

**Optimal Flows Over Time.** The *maximum flow-over-time* problem with time horizon $T$ is to maximize the amount of the flow sent from $s$ to $t$ by time $T$. Ford and Fulkerson [14] show that the maximum dynamic flow can be obtained in polynomial time by computing the static flow $\hat{f}$ which maximizes $(T+1)|\hat{f}| - \sum_e d_e \hat{f}_e$. Thus $\hat{f}$ is a minimum cost static flow with the cost of edges being the edge-delays. For a flow decomposition $\{\hat{f}_p\}_{p \in \mathcal{P}}$ of $\hat{f}$, the maximum dynamic flow sends flow at rate $|\hat{f}_p|$ along path $p$ from time 0 to $T - d_p$. Such a dynamic flow, obtained by repeating a static flow over time, is called a *temporally repeated flow*. For a maximum dynamic flow, we call the static flow repeated over time the *underlying static flow*.

The *quickest flow* problem for flow $M$ is to find the flow over time which minimizes the time taken to send $M$ units of flow from $s$ to $t$. The quickest flow problem can be solved by a binary search to find the minimum time $T$ for a maximum dynamic flow to route at least $M$ units of flow. Thus, the quickest flow problem can be solved by a temporally repeated flow.

The *earliest arrival flow* problem is to find a flow over time which maximizes the flow that arrives at the destination by time $\theta$, for every time $\theta$. An earliest arrival flow is also a maximum flow-over-time and a quickest flow, but the converse may not be true. Thus, the earliest arrival flow may not be a temporally repeated flow. For a single source and destination, earliest arrival flows exist [15].

**Temporal Routing Games.** The tuple $\Gamma = (G, s, t, c, d, c_0, M)$ forms an instance of the *temporal routing game*. Every player in this game controls infinitesimal flow. A player's cost is the time its flow arrives at the sink. A player's strategy is a path from $s$ to $t$. We assume an arbitrary ordering on the players which corresponds to the order in which their flow arrives at the source.

**Lemma 1** ([20])**.** *For any edge $e \in E$, the function $\theta + q_e(\theta)$ is monotonically increasing in $\theta$.*

By (9), the earliest time that flow entering an edge at time $\theta$ can exit the edge is $\theta + d_e + q_e(\theta)$. It follows from Lemma 1 that in a temporal routing game, flow does not wait on an edge unless the edge has a queue.

**Equilibrium Flow.** Informally, a flow over time is an equilibrium flow if every player minimizes its cost, given the strategies of the other players. To formalize this, for every vertex $v$ the label function $l_v(\theta)$ is the earliest time that flow starting from $s$ at time $\theta$ can reach $v$. Thus $l_s(\theta) = \theta$, and

$$l_v(\theta) = \min_{(u,v) \in E}\{l_u(\theta) + d_{uv} + q_{uv}(l_u(\theta))\}$$

From Lemma 1 and the definition of the label functions:

**Lemma 2.** *For each node $v \in V$, the function $l_v$ is monotonically increasing and continuous.*

For vertex $v$ and time $\theta$, define $l'_v(\theta) := \frac{\partial l_v(\theta)}{\partial \theta}$. Since $l_s(\theta) = \theta$, $l'_s(\theta) = 1$.

For a fixed $\theta$ and given the queues on the edges, the labels on all the vertices can be found in the following manner: $l_s(\theta) = \theta$ and $l_v(\theta) = \infty$ for $v \neq s$. Then $n - 1$ times, for each $e = (u, v) \in E$ set $l_v(\theta) = \min\{l_v(\theta), l_u(\theta) + d_{uv} + q_{uv}(l_u(\theta))\}$. The correctness of the labels obtained after $n - 1$ repetitions follows from the correctness of the Bellman-Ford algorithm.

The *shortest-path network* at time $\theta$, $G_\theta$, is the subgraph induced by the set of edges $E_\theta = \{(u,v) \in E : l_v(\theta) = l_u(\theta) + d_{uv} + q_{uv}(l_u(\theta))\}$. Flow is *sent over current shortest paths* if for every edge $(u,v) \in E$ and for all $\theta \in \mathbb{R}_+$, if $l_v(\theta) < l_u(\theta) + d_e + q_{uv}(l_u(\theta))$ then $f_e^+(l_u(\theta)) = 0$.

**Definition 3.** *Let $(G, s, t, c, d, c_0, M)$ be a temporal routing game. A flow over time $(f^+, f^-)$ is an equilibrium flow if*

*(i)* $\sum_{(s,v) \in E} f_{sv}^+(\theta) = \begin{cases} c_0 & \text{if } \theta \leq M/c_0 \\ 0 & \text{otherwise} \end{cases}$,

*(ii) flow is sent over current shortest paths, and*

*(iii) for every $e \in E$ and $\theta \geq 0$, if $q_e(\theta) > 0$, then $f_e^-(\theta + d_e) = c_e$.*

Every temporal routing game has an equilibrium [20].

For a temporal routing game $\Gamma$, $(f^+(\Gamma), f^-(\Gamma))$ is the equilibrium flow, and $EQ(\Gamma)$ is the completion time of the equilibrium flow. If the instance is clear from context, we simply use $(f^+, f^-)$ and $EQ$.

**Price of Anarchy.** In this paper we consider two separate objectives. In Section 4 and Section 6, our objective is to minimize the completion time. For this objective, the optimal flow is a quickest flow. Since a quickest flow can be represented as a temporally repeated static flow, we use $\hat{f}(\Gamma)$ to denote the underlying static flow for the quickest flow for a temporal routing game $\Gamma$, and $\hat{T}(\Gamma)$ to denote the completion time of the quickest flow. The time price of anarchy is then defined as $\max_\Gamma EQ(\Gamma)/\hat{T}(\Gamma)$. If the instance is clear, we use $\hat{f}$ to refer to the underlying static flow and $\hat{T}$ for the completion time.

We say the static flow underlying the quickest flow saturates every edge of the graph if for all $e \in E$, $\hat{f}_e = c_e$ and $\sum_v \hat{f}_{sv} = c_0$. We show that if this condition holds, then the price of anarchy is small.

Instead of flow entering the graph at the source $s$ at a constant rate $c_0$, another way to think of the model is that all flow is present at the same time at a node $s'$, and there is an initial edge $(s', s)$ of capacity $c_0$ and delay 0. Then the arrival time at $t$ for a player is also its delay. In Section 5, the objective is to minimize the total delay of a flow over time which routes a fixed amount of flow $M$ from the source to the destination. The total delay of a flow over time in a temporal routing instance is the sum of the arrival times at $t$ of the players. For a flow $(f^+, f^-)$ and instance $\Gamma$ with completion time $T$, the total delay $D((f^+, f^-)) = \int_0^T f_t^+(\theta)\theta d\theta$. In this case the earliest arrival flow is the optimal flow since it maximizes the flow at $t$ at every time $\theta$. For a temporal routing game $\Gamma$, let $(g^+(\Gamma), g^-(\Gamma))$ be the earliest arrival flow. The total delay price of anarchy is defined as $\max_\Gamma D(f^+(\Gamma), f^-(\Gamma))/D(g^+(\Gamma), g^-(\Gamma))$.

In the appendix, we give an example of a temporal routing game and its equilibrium flow.

## 3 The Structure of Equilibria

Equilibria in temporal routing games can be characterized in terms of static flows with certain properties, called *rate flows*. We use the properties of rate flows to obtain our bounds on the price of anarchy. In this section, we introduce some of these properties, as well as an algorithm for computing equilibria. Both rate flows and the algorithm we discuss are described in [20].

**Rate Flows.** For edge $e = (v, w) \in E$ and time $\theta \in \mathbb{R}_+$, define $x_e^+(\theta) := F_e^+(l_v(\theta))$ and $x_e^-(\theta) := F_e^-(l_w(\theta))$.

**Theorem 4** ([20]). *For a flow over time, flow is sent over current shortest paths if and only if for all edges and for all $\theta$, $x_e^+(\theta) = x_e^-(\theta)$.*

Let $x_e(\theta) := x_e^+(\theta)$. At equilibrium, it follows by integrating (4) over time and from Theorem 4 that for every $\theta \in \mathbb{R}_+$, $x_e(\theta)$ is a static flow in the uncapacitated network $G$. For $\theta$ such that $x(\theta)$ is differentiable,

$$\begin{aligned}\frac{\partial x_e(\theta)}{\partial \theta} &= f_e^+(l_v(\theta))l_v'(\theta) \\ &= f_e^-(l_w(\theta))l_w'(\theta)\,.\end{aligned} \qquad (10)$$

For any time $\theta$, the flow $x(\theta)$ given by $x_e(\theta)$ on every edge is called the static flow underlying the equilibrium flow. Let $x_e'(\theta) := \frac{dx_e(\theta)}{d\theta}$ where the differential exists. The following theorem describes some properties of $x'(\theta)$. For $\theta \in \mathbb{R}_+$, define $E_1(\theta) := \{(v,w) \in E : q_{vw}(l_v(\theta)) > 0\}$ as the set of edges which have positive queues on them at time $\theta$.

**Theorem 5** ([20])**.** *For an equilibrium flow in a temporal routing game $\Gamma = (G, s, t, c, d, c_0)$ let $\theta \geq 0$ be such that $x_e'(\theta)$ and $l_v'(\theta)$ exist for all $v \in V$, $e \in E$. Then $(x_e'(\theta))_{e \in G_\theta}$ is a static flow of value $c_0$ in the uncapacitated graph. Further, the static flow $(x_e'(\theta))_{e \in G_\theta}$ satisfies*

$$\begin{aligned}l_w'(\theta) &\leq l_v'(\theta), & \forall (v,w) \in E(G_\theta) \setminus E_1(\theta) \text{ with } x_{vw}'(\theta) = 0\,, \\ l_w'(\theta) &= \max\left\{l_v'(\theta), \frac{x_{vw}'(\theta)}{c_{vw}}\right\} & \forall (v,w) \in E(G_\theta) \setminus E_1(\theta) \text{ with } x_{vw}' > 0\,, \\ l_w'(\theta) &= \frac{x_{vw}'(\theta)}{c_{vw}} & \forall (v,w) \in E_1(\theta)\,.\end{aligned}$$

The static flow $(x_e'(\theta))_{e \in G_\theta}$ is called a *rate flow*. By (10), $x_{vw}'(\theta)$ exists iff $l_v'(\theta)$ and $l_w'(\theta)$ exist. If at times $\theta$ and $\theta' \in \mathbb{R}_+$ the shortest-path networks and the set of edges with positive queues are the same, i.e., $G_\theta = G_{\theta'}$ and $E_1(\theta) = E_1(\theta')$, then the rate flow $x'(\theta)$ and $l_v'(\theta)$ satisfy the conditions of Theorem 5 at time $\theta'$ as well.

**Computing Equilibria.** In [20], the authors describe an algorithm to compute equilibrium flow. The algorithm divides the time from $\theta = 0$ to $\theta = M/c_0$ into a number of phases, with phases divided by events. Note that $\theta = M/c_0$ is the time the last flow leaves the source; the time this flow reaches the sink is the completion time. An event can be of two kinds. A *queue-event* occurs at time $\theta$ if for some edge $e = (u,v)$, the queue decreases to zero at time $l_u(\theta)$. That is, the queueing delay $q_e(l_u(\theta)) = 0$ and for some $\delta > 0$ and every $0 < \epsilon \leq \delta$, $q_e(l_u(\theta - \epsilon)) > 0$. A *path-event* occurs at time $\theta$ if some edge $e = (u,v)$ enters the shortest-path network at time $\theta$, i.e., $l_v(\theta) = l_u(\theta) + d_e + q_e(l_u(\theta))$ and for some $\delta > 0$ and every $0 < \epsilon \leq \delta$, $l_v(\theta) > l_u(\theta) + d_e + q_e(l_u(\theta))$. Queue events and path events are collectively termed *events*. Note that in determining the time an event occurs, we are using the source as a frame of reference. While the event actually occurs at a later time $\theta'$, we say an event occurs at time $\theta$ if flow leaving the source at time $\theta$ reaches the tail of the edge in the queue-event or path-event at time $\theta'$.

For a given instance, we order the events occurring in an equilibrium flow by the time of the occurrence (using the source as the frame of reference) and index them, starting from 0 to $r$. The event $r$ is a special event, corresponding to the last flow leaving the source, thus the equilibrium flow ends at event $r$. We define $\theta_i$ as the time event $i$ occurs, and $\tau_i := l_t(\theta_i)$. Thus, $\theta_r = M/c_0$ and $\tau_r = EQ$.

A *phase* is the time interval between two events. Phase $i$ as the time between events $i-1$ and $i$. Thus, time $\theta$ is in phase $i$ if $\theta_{i-1} < \theta < \theta_i$. We exclude the event times $\theta_i$ since the vertex labels are $l_v(\theta)$ and the static flow $x(\theta)$ are not differentiable at these times. Within a phase, the shortest-path network and the set of edges with queues on them remain constant. Hence the rate flow $x'(\theta)$ and the rate of change of vertex labels $l'_v(\theta)$ exist and are fixed for $\theta$ within a phase. The first phase, phase 1, is the time between $\theta_0 = 0$ and $\theta_1$. Thus, for a phase $i$, we define the following notation:

- $G_i$ denotes the shortest-path network in phase $i$.
- $\tilde{c}_i$ is the capacity of the shortest path network in phase $i$.
- $\triangle_i$ is the change in capacity of the shortest path network when event $i$ occurs, thus $\triangle_i = \tilde{c}_i - \tilde{c}_{i-1}$. We define $\tilde{c}_0 := 0$.

Note that $\triangle_i = 0$ if event $i-1$ is a queue event, or $i-1$ is a path event but the capacity of shortest path network does not change; this could happen if a minimum cut is unaffected by the edge added to the shortest path network in a path event.

Since the rate of change of vertex labels $l'_v(\theta)$ is fixed for $\theta$ within a phase, the vertex labels $l_v(\theta)$ are fixed linear functions of $\theta$ within a phase; and thus within a phase $l'_v$ is well-defined for all $v$. Thus, we can define:

- $l_v^{i\,\prime} := l'_v(\theta)$ for any time $\theta$ in phase $i$.
- The set $E_1^i$ is defined as $\{e = (v,w) : q_e(l_v(\theta_i)) > 0\}$.
- We use $x'_i$ to denote the rate flow in phase $i$.

For the notation above, if the phase is clear from context, for simplicity we omit the phase. Thus the rate flow would be denoted by $x'$.

For edge $(v,w)$ in the shortest path network at time $\theta$, $q'_{vw}(\theta) := \frac{\partial q_{vw}(l_v(\theta))}{\partial \theta}$. Since $l_w(\theta) = l_v(\theta) + d_{vw} + q_{vw}(l_v(\theta))$, $q'_{vw}(\theta) = l'_w(\theta) - l'_v(\theta)$. Since the rate of change of the vertex labels is constant, we define for phase $i$:

- If edge $e = (v,w)$ is in the shortest-path network in phase $i$, define $q_{vw}^{i\,\prime} := l_w^{i\,\prime} - l_v^{i\,\prime}$, otherwise $q_e^{i\,\prime} := 0$.
- For an $s$-$t$ path $p$, we abuse notation slightly to define $q_p^{i\,\prime} := \sum_{e \in p} q_e^{i\,\prime}$.

## 4 A Stackelberg Strategy for the Time Price of Anarchy

In Section 6, we prove our main technical result:

**Theorem 6.** *For a temporal routing game where the static flow underlying the quickest flow saturates every edge of the graph, the time price of anarchy is $e/(e-1)$.*

In general instances of the temporal routing game where the rate of equilibrium flow may exceed the optimal flow on some edges, a bound on the time price of anarchy is unknown. However, in any temporal routing game, we show how to use Theorem 6 to obtain a simple Stackelberg strategy to enforce a bound of $e/(e-1)$ on the price of anarchy.

**Theorem 7.** *For a temporal routing game, let $\hat{T}$ be the time taken by the quickest flow to route all flow to the sink. There exists a polynomial-time computable Stackelberg strategy to enforce an equilibrium flow that routes all flow at the source to the sink in time at most $\hat{T} \times e/(e-1)$.*

*Proof.* For a temporal routing game $\Gamma = (G, s, t, c, d, c_0, M)$, let $\hat{f}$ be the static flow underlying the quickest flow. This can be computed in polynomial time by conducting a binary search to find the minimum time $T$ such that the maximum dynamic flow with time horizon $T$ gets at least $M$ flow to the sink.

The Stackelberg strategy is then as follows. The network manager, acting as the leader, reduces the capacity on each edge so that the new capacities $c'$ are the value of the static flow on each edge in $\Gamma$: $c'_e = \hat{f}_e$. It is easy to see that the quickest flow remains unchanged; further on each edge with the modified capacities, $\hat{f}_e$ saturates every edge. By Theorem 6, the price of anarchy is now bounded by $e/(e-1)$. □

Thus in any instance of the temporal routing game, by reducing the capacity of edges, the completion time of equilibrium flow can be bounded by $e/(e-1)$ times the completion time of the optimal flow.

## 5 The Total Delay Price of Anarchy

We now obtain bounds on the total delay price of anarchy of temporal routing games. The total delay price of anarchy is the maximum over all instances, of the ratio of total delay of the equilibrium flow to the minimum total delay. Since the cost of a player is the time it arrives at the sink, for a flow over time $(f^+, f^-)$ with completion time $T$ the total delay $D((f^+, f^-)) = \int_0^T \theta f_t^+(\theta) d\theta$.

We first show that in temporal routing games which satisfy the same assumption as in Theorem 6, the total delay price of anarchy is bounded by a small constant.

**Theorem 8.** *For a temporal routing game where the static flow underlying the quickest flow saturates every edge of the graph, the total delay price of anarchy is $2e/(e-1)$.*

The following lemma gives a lower bound on the total delay of the earliest arrival flow. The proof, and all missing proofs, are given in the appendix.

**Lemma 9.** *The total delay of the earliest arrival flow $(g^+, g^-)$ with completion time $T$ in an instance $\Gamma$ is at least $MT/2$.*

*Proof of Theorem 8.* Let $EQ$ denote the completion time of the equilibrium flow. Then by Theorem 6, $EQ \leq Te/(e-1)$. The total delay of equilibrium flow $(f^+, f^-)$ is bounded by

$$
\begin{aligned}
D((f^+, f^-)) &= \int_0^{Te/e-1} \theta f_t^+(\theta) d\theta \\
&\leq T \frac{e}{e-1} \int_0^{Te/(e-1)} f_t^+(\theta) d\theta \\
&\leq MT \frac{e}{e-1}
\end{aligned}
\qquad (11)
$$

The result now follows from (11) and Lemma 9. □

Similar to the proof of Theorem 7, Theorem 8 can be used to give a Stackelberg strategy for enforcing a bound of $2e/(e-1)$ on the total delay price of anarchy in any general instance.

**Theorem 10.** *For a temporal routing game, there exists a polynomial-time computable Stackelberg strategy to enforce an equilibrium flow with total delay at most $2e/(e-1)$ times that of the earliest arrival flow in the unmodified instance.*

# 6 The Time Price of Anarchy

In this section, we prove Theorem 6. We assume that on every edge, $\hat{f}_e = c_e$. For a path decomposition $\{\hat{f}_p\}_{p \in \mathcal{P}}$ of $\hat{f}$ along paths $p \in \mathcal{P}$, by our assumption, $\sum_{p \in \mathcal{P}} \hat{f}_p = c_0$.

Conceptually, we show that for an instance $\Gamma$ of the temporal routing game, the ratio of $EQ$ to $\hat{T}$ is worst if every event either occurs at time 0, or occurs at a fixed time $\mu$. Thus, we modify an instance $\Gamma$ to obtain an instance $\Gamma'$ where every event either occurs at time 0 or time $\mu$. We then obtain a bound on $\frac{EQ}{\hat{T}}$ in this simpler instance.

This conceptual view is simplified; it may not always be possible to preserve the events if we insist on every event occuring at either time 0 or time $\mu$. However, we show a bound on $\frac{EQ}{\hat{T}}$ can be obtained in $\Gamma$ by following the same steps analytically:

**Step 1:** For any path $p$, get a lower bound on $d_p$ in terms of $\{\theta_i\}_{i=0}^r$ and the queues on the path $p$ (Lemma 14).

**Step 2:** Use the bound in step 1 to obtain an upper bound on $\frac{EQ}{\hat{T}}$ in terms of the event times $\{\theta_i\}_{i=0}^r$ and the queues on the edges (Lemma 15 and Corollary 16).

**Step 3:** Show that there is some event $k \leq r$ so that if all the events before and including $k$ happen at time 0, and all events after $k$ occur at the same time, then the upper bound on $\frac{EQ}{\hat{T}}$ in this modified instance also bounds $\frac{EQ}{\hat{T}}$ in the original instance (Lemma 17).

**Step 4:** Evaluate the upper bound on $\frac{EQ}{\hat{T}}$ for this modified, simpler instance (Lemma 18 and Theorem 6).

Our first step is to show a relation between the label on $t$ and the rate of flow into $t$.

**Lemma 11.** *Let $(f^+, f^-)$ be the equilibrium flow for a temporal routing game $\Gamma$ with inflow $c_0$ and corresponding labels $l$. Then $l'_t(\theta) = \dfrac{c_0}{f_t^+(l_t(\theta))}$ for $\theta \in \mathbb{R}_+$.*

We first show Lemma 11 for a path in the graph, and then use path decompositions of $x'(\theta)$ in conjunction with (10) to get the result.

**Lemma 12.** *Let $p = (s, v_1, v_2, \ldots, v_k)$ be a path in $G_\theta$. If for every pair of consecutive edges $(u, v)$, $(v, w)$ in $p$, $f_{uv}^-(l_v(\theta)) = f_{vw}^+(l_v(\theta))$, then $l'_{v_k}(\theta) = \dfrac{f_{sv_1}^+(l_s(\theta))}{f_{v_{k-1}v_k}^-(l_{v_k}(\theta))}$.*

*Proof of Lemma 11.* Let $x'$ and $l'_v$ be the rate flow and rate of change of labels at time $\theta$. Let $\{x'_p\}_{p \in \mathcal{P}}$ be a path decomposition of $x'$ where $\mathcal{P}$ is the set of paths with positive flow. Instead of the graph $G = (V, E)$, we consider the equilibrium flow in a graph $\bar{G} = (V, \bar{E})$ with $\bar{E} = E \times \mathcal{P}$. Every edge $a \in \bar{E}$ corresponds to a pair $(e, p)$ with $e \in E$ and $p \in \mathcal{P}$, with capacity $c_a = c_e \frac{x'_p}{x'_e}$ and delay $d_a = d_e$. We obtain an equilibrium flow in $\bar{G}$ and show that the labels on the vertices at any time $\phi$ are the same in $G$ and $\bar{G}$.

Define a modified flow over time $(\bar{f}^+, \bar{f}^-)$ in $\bar{G}$ as follows: $\bar{f}_a^+(\phi) = f_e^+(\phi)\frac{c_a}{c_e}$ and $\bar{f}_a^-(\phi) = f_e^-(\phi)\frac{c_a}{c_e}$. Then the cumulative flow $\bar{F}_a^+(\phi) := \int_0^\phi \bar{f}_a^+(\theta) = F_e^+ \frac{c_a}{c_e}$ and similarly $\bar{F}_a^-(\phi) := \int_0^\phi \bar{f}_a^-(\phi) = F_e^- \frac{c_a}{c_e}$. Thus, $q_a(\phi) = q_e(\phi)$ via (9). Since $d_a = d_e$ and $l_s(\phi) = \phi$, it follows that the labels on the vertices in $\bar{G}$ for the flow over time $(\bar{f}^+, \bar{f}^-)$ are equal to the labels on the corresponding vertices in $G$ for the equilibrium flow $(f^+, f^-)$, for every time $\phi \in \mathbb{R}_+$. It is easy to verify the conditions for equilibrium flow in Definition 3 for $(\bar{f}^+, \bar{f}^-)$ in $\bar{G}$.

We show that at time $\theta$, $l'_t = \dfrac{c_0}{\bar{f}_t^+(l_t(\theta))}$ in $H$. Since the node labels are the same for $(f^+, f^-)$ and $(\bar{f}^+, \bar{f}^-)$, and $f_v^+(\phi) = \bar{f}_v^+(\phi)$ for every vertex $v$ and time $\phi \in \mathbb{R}_+$, this proves the lemma. Since $(\bar{f}^+, \bar{f}^-)$ is an equilibrium flow in $\bar{G}$, the flow $y'$ with $y'_e = x'_e \frac{c_a}{c_e}$ is a rate flow in graph $\bar{G}$ at time $\theta$. Consider a path $p \in \mathcal{P}$; there is a corresponding $s$-$t$ path $q$ in $\bar{G}$ consisting of all the edges which correspond to path $p$. By our construction, on every edge $a$ of path $q$, $y'_a = x'_p$.

On consecutive edges $(u, v), (v, w)$ in $q$, $y'_{uv} = l'_v \bar{f}_{uv}^-(l_v(\theta))$ and $y'_{vw} = l'_v \bar{f}_{vw}^+(l_v(\theta))$. Since $y'_{uv} = y'_{vw}$, it follows that $\bar{f}_{uv}^-(l_v(\theta)) = \bar{f}_{vw}^+(l_v(\theta))$. Then by Lemma 12, $l'_t \bar{f}_{v_k t}^-(l_t(\theta)) = \bar{f}_{sv_1}^+(l_s(\theta))$. Since this is true for all paths $p \in \mathcal{P}$, we can sum over all these paths to obtain $l'_t(\theta)\bar{f}^+_t(l_t(\theta)) = \bar{f}^+_s(\theta) = c_0$. Thus in graph $\bar{G}$, $l'_t = \frac{c_0}{\bar{f}^+_t(l_t(\theta))}$. □

**Corollary 13.** *For any events $i$, $i-1$, $\tau_i - \tau_{i-1} = \frac{c_0}{\tilde{c}_i}(\theta_i - \theta_{i-1})$.*

*Proof.* By definition, $\tau_i - \tau_{i-1} = l_t(\theta_i) - l_t(\theta_{i-1})$ and $f^+_t(l_t(\theta)) = \tilde{c}_i$ in phase $i$. Thus $\tau_i - \tau_{i-1} = \int_{\theta_{i-1}}^{\theta_i} l'_t(\phi) d\phi = \int_{\theta_{i-1}}^{\theta_i} \frac{c_0}{\tilde{c}_i} d\phi = \frac{c_0}{\tilde{c}_i}(\theta_i - \theta_{i-1})$. □

We use Corollary 13 to bound $d_p$ in terms of $\{\tau_i\}_{i=0}^r$.

**Lemma 14.** *For any s-t path $p$, $d_p \geq \tau_r - \sum_{i=1}^r \left(1 + q_p^{i'}\right) \frac{\tilde{c}_i}{c_0}(\tau_i - \tau_{i-1})$.*

*Proof.* By definition of shortest path network, for any time $\theta$ and for any edge $e = (v,w)$ in the shortest path network $G_\theta$, $d_e = l_w(\theta) - l_v(\theta) - q_e(l_v(\theta))$. For any vertex $v$, $l_v(\theta_r) = \sum_{i=1}^r l_v^{i'}(\theta_i - \theta_{i-1}) + l_v(\theta_0)$, and similarly $q_e(l_v(\theta_r)) = \sum_{i=1}^r q_e^{i'}(\theta_i - \theta_{i-1})$. Hence, for any edge in the shortest path network at time $\theta_r$, $d_e = l_w(\theta_0) - l_v(\theta_0) + \sum_{i=1}^r \left(l_w^{i'} - l_v^{i'} - q_e^{i'}\right)(\theta_i - \theta_{i-1})$. For edges not in the shortest path network at $\theta_r$, $d_e > l_w(\theta_r) - l_v(\theta_r) = l_w(\theta_0) - l_v(\theta_0) + \sum_{i=1}^r \left(l_w^{i'} - l_v^{i'} - q_e^{i'}\right)(\theta_i - \theta_{i-1})$. Summing over all edges in path $p$ yields

$$d_p \geq \tau_0 + \sum_{i=1}^r \left(l_t^{i'} - l_s^{i'} - q_p^{i'}\right)(\theta_i - \theta_{i-1}).$$

Substituting in from Corollary 13 and from Lemma 11, and since $l'_s(\theta) = 1$,

$$d_p \geq \tau_0 + \sum_{i=1}^r \frac{\tilde{c}_i}{c_0}\left(\frac{c_0}{\tilde{c}_i} - 1 - q_p^{i'}\right)(\tau_i - \tau_{i-1}).$$

Simplifying yields the desired result. □

**Lemma 15.** *For a temporal routing game with $\sum_{p \in \mathcal{P}} \hat{f}_p = c_0$, the completion times of the optimal flow and equilibrium flow are related as*

$$\frac{\hat{T}}{EQ} = \frac{\tilde{c}_r}{c_0} - \frac{1}{c_0 EQ}\left(\sum_{i=0}^{r-1} \tau_i \triangle_{i+1} - \sum_{p \in \mathcal{P}} \hat{f}_p d_p\right). \quad (12)$$

The proof is based on the flow arrival rate at $t$ for the equilibrium and optimal flows. For the temporal routing game in Figure 2, these arrival rates are plotted in Figure 3 and Figure 4.

*Proof.* Consider the arrival rate at the sink for the optimal flow. For any path $p \in \mathcal{P}$, the rate of flow arriving at the sink increases by $\hat{f}_p$ at time $d_p$. The total flow arriving at the sink by time $\theta$ is the area under this curve up to time $\theta$. Thus, $M = c_0 \hat{T} - \sum_p \hat{f}_p d_p$, and similarly for the equilibrium flow, $M = \tilde{c}_r EQ - \sum_{i=0}^{r-1} \tau_i \triangle_{i+1}$. Equating these yields

$$c_0 \hat{T} = \tilde{c}_r EQ + \sum_{p \in \mathcal{P}} \hat{f}_p d_p - \sum_{i=0}^{r-1} \tau_i \triangle_{i+1},$$

and dividing both sides by $c_0 EQ$ yields the desired equality. □

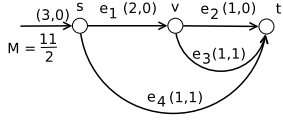 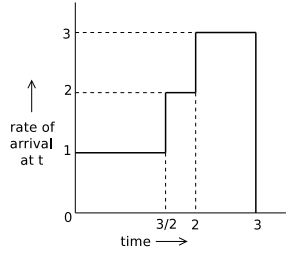 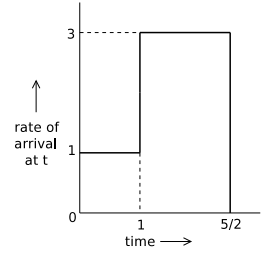

Figure 2: An instance of a temporal routing game

Figure 3: Arrival rate at $t$ for equilibrium flow

Figure 4: Arrival rate at $t$ for optimal flow

Using the lower bound in Lemma 14 and defining $\lambda_r := \tilde{c}_r - c_0 + \frac{\tilde{c}_r}{c_0} \sum_p \hat{f}_p q_p^{r'}$, $\lambda_0 := -\frac{\tilde{c}_1}{c_0} \sum_p \hat{f}_p q_p^{1'}$ and for $1 \leq i \leq r-1$, $\lambda_i := \frac{1}{c_0} \sum_p \hat{f}_p \left( \tilde{c}_i q_p^{i'} - \tilde{c}_{i+1} q_p^{i+1'} \right)$,

**Corollary 16.** *For a temporal routing game with $\sum_{p \in \mathcal{P}} \hat{f}_p = c_0$,*

$$\frac{\hat{T}}{EQ} \geq \frac{\tilde{c}_r}{c_0} - \frac{1}{c_0 \, EQ} \sum_{i=0}^{r} \lambda_i \tau_i \,.$$

Lemma 17 is used to partition the events into two sets, with events in the first set occurring at time $\theta = 0$ and events in the second set occurring at time $\theta_r$:

**Lemma 17.** *For $0 \leq i \leq r$ and $\lambda_i, y_i \in \mathbb{R}$, if $0 \leq y_0 \leq y_1 \leq \cdots \leq y_r$, then $\sum_{i=0}^{r} \lambda_i y_i \leq y_r \max_k \sum_{i=k}^{r} \lambda_i$.*

By Lemma 17, $\exists k \leq r : \sum_{i=0}^{r} \lambda_i \tau_i \leq \tau_r \sum_{i \geq k} \lambda_i$. Then since $\tau_r = EQ$, substituting in Corollary 16,

$$\frac{\hat{T}}{EQ} \geq \frac{\tilde{c}_r}{c_0} - \frac{1}{c_0} \sum_{i \geq k} \lambda_i \,. \tag{13}$$

Evaluating $\sum_{i \geq k} \lambda_i$, we obtain $\sum_{i \geq k} \lambda_i = \tilde{c}_r - c_0$ if $k = 0$ and $\sum_{i \geq k} \lambda_i = \tilde{c}_r - c_0 + \frac{\tilde{c}_k}{c_0} \sum_p \hat{f}_p q_p^{k'}$ if $k > 0$. If $k = 0$, then (13) becomes $\frac{\hat{T}}{EQ} \geq \frac{\tilde{c}_r}{c_0} - \frac{1}{c_0}(\tilde{c}_r - c_0) = 1$ and hence in this case, $\hat{T} = EQ$. If $k > 0$,

$$\begin{aligned}
\frac{\hat{T}}{EQ} &\geq \frac{\tilde{c}_r}{c_0} - \frac{1}{c_0} \left( \tilde{c}_r - c_0 + \frac{\tilde{c}_k}{c_0} \sum_p \hat{f}_p q_p^{k'} \right) \\
&= 1 - \frac{\tilde{c}_k}{(c_0)^2} \sum_p \hat{f}_p q_p^{k'} \\
&= 1 - \frac{\tilde{c}_k}{(c_0)^2} \sum_e \hat{f}_e q_e^{k'} \\
&= 1 - \frac{\tilde{c}_k}{(c_0)^2} \sum_e c_e q_e^{k'} \,, \tag{14}
\end{aligned}$$

since by assumption, $c_e = \hat{f}_e$ on every edge.

**Lemma 18.** *In any phase $k$ of the equilibrium flow, $\sum_e c_e q_e^{k'} \leq c_0 \ln \frac{c_0}{\tilde{c}_k}$.*

*Proof Sketch.* By the conditions of Theorem 5 and the definition of $q_e^{k'}$, $c_e q_e^{k'} = x_e^{k'}(1 - \frac{l_v^{k'}}{l_w^{k'}})$ and hence $\sum_e c_e q_e^{k'} = \sum_{e=(v,w)} x_e^{k'} \left(1 - \frac{l_v^{k'}}{l_w^{k'}}\right) = \sum_{p \in \mathcal{P}} x_p^{k'} \sum_{e=(v,w) \in p} \left(1 - \frac{l_v^{k'}}{l_w^{k'}}\right)$ for a path decomposition $\{x_p^{k'}\}_{p \in \mathcal{P}}$ of $x^{k'}$. We then show that for any $s$-$t$ path $p$, $\sum_{e=(v,w) \in p} \left(1 - \frac{l_v^{k'}}{l_w^{k'}}\right) \leq \ln l_t^{k'} = \ln \frac{c_0}{\tilde{c}_k}$ by Lemma 11. Since $\sum_{p \in \mathcal{P}} x_p^{k'} = c_0$, the result follows.

*Proof of Theorem 6.* Let $w = \tilde{c}_k/c_0$. Then from (14) and Lemma 18, $\frac{\hat{T}}{EQ} \geq 1 - w \ln \frac{1}{w}$. Let $z = w \ln \frac{1}{w}$, then $z$ is maximized when $w = 1/e$. Hence, $\frac{\hat{T}}{EQ} \geq 1 - 1/e = \frac{e-1}{e}$. Observing that $\frac{\hat{T}}{EQ}$ is the inverse of the price of anarchy, completes the proof. □

Note that we did not use any properties of the optimal flow over time in our proof. Instead of the optimal flow over time, we could obtain the same results for any temporally repeated static flow, with $\hat{f}$ being the static flow.

## A  An Example

We present a brief example to demonstrate an equilibrium flow over time. For the temporal routing game in Figure 2, $M = 5.5$, $c_0 = 3$, and each edge is marked $(c_e, d_e)$. In the example, the equilibrium flow is calculated as follows. Given the shortest-path network and the set of edges with queues at the start of a phase, we choose the rate low $x^{i'}$ so that there exist labels $l_v^{i'}$ which satisfy the conditions of Theorem 5. Given the rate flow $x^{i'}$ and the vertex labels $l_v^{i'}$ within a phase, the equilibrium flow within the phase can be obtained by (10). A phase ends when either a path event or queue event occurs, or there is no more flow at the source.

Phase 1 starts at $\theta_0 = 0$ and on every vertex $v$ the label $l_v(\theta_0) = 0$. In this phase, edges $e_1$ and $e_2$ are in the shortest path network. Thus $x'_e = c_0 = 3$ on both of these edges. We use the conditions in Theorem 5 to obtain the rate of change of labels, given by $l'_s = 1$, $l'_v = 3/2$, and $l'_t = 3$. Hence $q'_{e_1} = 1/2$, and $q'_{e_2} = 3/2$. Phase 0 ends when edge $e_4$ enters the shortest-path network, at $\theta_1$ such that $l_t(\theta_1) - l_s(\theta_1) = 1$, yielding $\theta_1 = 1/2$. The equilibrium flow over time can be calculated by (10), to obtain $f_{e_1}^+(l_s(\theta)) = 3$, $f_{e_1}^-(l_v(\theta)) = f_{e_2}^+(l_v(\theta)) = 2$, $f_{e_2}^-(l_t(\theta)) = 1$ for $\theta \in (\theta_0, \theta_1)$.

Phase 2 starts at $\theta_1 = 1/2$, and edges $e_1$, $e_2$ and $e_4$ are in the shortest path network. The rate flow and the rate of change of vertex labels are again obtained by the conditions in Theorem 5, so that $x'_{e_1} = x'_{e_2} = 3/2$, and $x'_{e_4} = 3/2$. Note that in this phase edges $e_1$ and $e_2$ had queues on them at the start of the phase, hence $E_1^1 = \{e_1, e_2\}$. Then $l'_v = 3/4$, and $l'_t = 3/2$ and for the queues, $q'_{e_1} = -1/4$, $q'_{e_2} = 3/4$, and $q'_{e_4} = 1/2$. For $\theta \in (\theta_1, \theta_2)$ the equilibrium flow $f_{e_1}^+(l_s(\theta)) = 3/2$, $f_{e_1}^-(l_v(\theta)) = f_{e_2}^+(l_v(\theta)) = 2$, $f_{e_2}^-(l_t(\theta)) = 1$, and $f_{e_4}^+(l_s(\theta)) = 3/2$, $f_{e_4}^-(l_t(\theta)) = 1$. Phase 2 ends when edge $e_3$ enters the shortest path network at $\theta_2$ so that $l_t(\theta_2) - l_v(\theta_2) = 1$, yielding $\theta_2 = 5/6$.

In Phase 3 all the edges are in the shortest path network. The queues do not change on any edge, and this phase continues until the completion time. In this phase $x'_{e_1} = 2$, $x'_{e_2} = x'_{e_3} = 1$, and $x'_{e_4} = 1$. On all vertices, the labels $l'_v = 1$.

The rate of flow arrival at the sink as a function of time for this example is plotted in Figure 3.

## B  Proofs from Section 4

*Proof of Lemma 9.* The total delay of the earliest arrival flow is

$$D((g^+, g^-)) = \int_0^T \theta\, g_t^+(\theta) d\theta$$

$$= \int_0^{T/2} \theta\, g_t^+(\theta) d\theta - \int_T^{T/2} \theta\, g_t^+(\theta) d\theta$$

$$= \int_0^{T/2} \theta\, g_t^+(\theta) d\theta + \int_0^{T/2} (T - \phi)\, g_t^+(T - \phi) d\phi \qquad (15)$$

where the last inequality is obtained by substituting $\phi = T - \theta$. Since $g_t^+(\theta)$ is an increasing function of $\theta$ for $\theta \leq T$, $g_t^+(T - \theta) \geq g_t^+(\theta)$ for $\theta \leq T/2$, and hence

$$\theta\, g_t^+(\theta) + (T-\theta)\, g_t^+(T-\theta) \geq T/2\,(g_t^+(\theta) + g_t^+(T-\theta)).$$

Replacing in (15),

$$D((g^+, g^-)) \geq \int_0^{T/2} T/2\,(g_t^+(\theta) + g_t^+(T-\theta))d\theta$$
$$= MT/2$$

since $\int_0^{T/2}(g_t^+(\theta) + g_t^+(T-\theta))d\theta$ is the total flow arriving at $t$. □

## C  Proofs from Section 5

*Proof of Theorem 10.* The strategy for enforcing an equilibrium flow over time of total delay at most $2e/(e-1)$ times the total delay of the earliest arrival flow in general graphs is now exactly the algorithm described in the proof of Theorem 7. In particular, for an instance $\Gamma$ instead of looking at the earliest-arrival flow which may have an exponential-sized description, we only need to consider the quickest flow which can be computed in polynomial time. The network manager, as the leader, then reduces the capacities of every edge to equal the flow on the edge in the static flow underlying the quickest flow. By Theorem 8, the equilibrium flow obtained following the action of the leader has total delay at most $2e/(e-1)$ times the minimum total delay of any flow in the original instance. □

## D  Proofs from Section 6

*Proof of 12.* Let $x'$ and $l'$ denote the rate flow and rate of change of vertex labels at time $\theta$. For consecutive edges $(u,v)$, $(v,w)$ in $p$, $x'_{uv} = l'_v f^-_{uv}(l_v(\theta))$ and $x'_{vw} = l'_v f^+_{vw}(l_v(\theta))$ by (10). From the conditions of the lemma, it follows that $x'_{uv} = x'_{vw}$, and hence $x'_e = f^+_{sv_1}(\theta)$ for every edge in the path.

The proof is by induction on the size of the path. For the base case, $p$ is a single edge $e = (s, v)$. If $e$ has a queue on it at time $\theta$, then $f_e^-(l_v(\theta)) = c_e$ and $l'_v = x'_e/c_e = f_{sv}^+(\theta)/f_e^-(l_v(\theta))$, and hence the lemma is true. If $e$ does not have a queue, $l_v(\theta) = l_s(\theta) + d_e$, and hence $l'_v = l'_s = 1$. Since if $e$ does not have a queue $f_e^-(l_v(\theta)) = f_e^+(\theta)$, the lemma is true for the base case.

Let $p = (s, v_1, v_2, \ldots, v_k)$ be a path in $G_\theta$ of length $k$ and let $e = (v_{k-1}, v_k)$. If $e$ has a queue on it at time $l_{v_{k-1}}(\theta)$, then $f_e^-(l_{v_k}(\theta)) = c_e$ and $l'_{v_k} = x'_e/c_e = f^+_{sv_1}(\theta)/f_e^-(l_{v_k}(\theta))$. If edge $e$ does not have a queue on it, $l'_{v_k} = l'_{v_{k-1}} = \frac{f^+_{sv_1}(l_s(\theta))}{f^-_{v_{k-2}v_{k-1}}(l_{v_{k-1}}(\theta))}$ where the second equality follows from the inductive hypothesis. By the conditions of the lemma, $f^-_{v_{k-2}v_{k-1}}(l_{v_{k-1}}(\theta)) = f_e^+(l_{v_{k-1}}(\theta))$. Since edge $e$ does not have a queue on it, $f_e^+(l_{v_{k-1}}(\theta)) = f_e^-(l_{v_k}(\theta))$, and hence $l'_{v_k} = \frac{f^+_{sv_1}(l_s(\theta))}{f_e^-(l_{v_k}(\theta))}$. □

*Proof of Corollary 16.* From Lemma 14 and observing that $\sum_p \hat{f}_p = c_0$, and by rearranging terms,

$$\sum_{p\in\mathcal{P}} \hat{f}_p d_p \geq \sum_{p\in\mathcal{P}} \hat{f}_p \left[ \tau_r - \sum_{i=1}^{r} \left(1 + q_p^{i'}\right) \frac{\tilde{c}_i}{c_0} (\tau_i - \tau_{i-1}) \right]$$

$$= c_0 \tau_r - \sum_{i=1}^{r} (\tau_i - \tau_{i-1}) \frac{\tilde{c}_i}{c_0} \sum_{p\in\mathcal{P}} \hat{f}_p (1 + q_p^{i'})$$

$$= c_0 \tau_r - \sum_{i=1}^{r} \tilde{c}_i (\tau_i - \tau_{i-1}) - \sum_{i=1}^{r} \frac{\tilde{c}_i}{c_0} (\tau_i - \tau_{i-1}) \sum_{p\in\mathcal{P}} \hat{f}_p q_p^{i'}$$

$$= c_0 \tau_r - \tilde{c}_r \tau_r + \sum_{i=0}^{r-1} \tau_i \triangle_{i+1} + \tau_0 \frac{\tilde{c}_1}{c_0} \sum_{p\in\mathcal{P}} \hat{f}_p q_p^{1'}$$

$$- \tau_r \frac{\tilde{c}_r}{c_0} \sum_{p\in\mathcal{P}} \hat{f}_p q_p^{r'} + \sum_{i=1}^{r-1} \frac{\tau_i}{c_0} \sum_{p\in\mathcal{P}} \hat{f}_p \left( \tilde{c}_{i+1} q_p^{i+1'} - \tilde{c}_i q_p^{i'} \right),$$

and hence,

$$\sum_{p\in\mathcal{P}} \hat{f}_p d_p - \sum_{i=0}^{r-1} \tau_i \triangle_{i+1} \geq c_0 \tau_r - \tilde{c}_r \tau_r + \tau_0 \frac{\tilde{c}_1}{c_0} \sum_{p\in\mathcal{P}} \hat{f}_p q_p^{1'} - \tau_r \frac{\tilde{c}_r}{c_0} \sum_{p\in\mathcal{P}} \hat{f}_p q_p^{r'}$$

$$+ \sum_{i=1}^{r-1} \frac{\tau_i}{c_0} \sum_{p\in\mathcal{P}} \hat{f}_p \left( \tilde{c}_{i+1} q_p^{i+1'} - \tilde{c}_i q_p^{i'} \right).$$

Substituting $\lambda_r = \tilde{c}_r - c_0 + \frac{\tilde{c}_r}{c_0} \sum_p \hat{f}_p q_p^{r'}$, $\lambda_0 = -\frac{\tilde{c}_1}{c_0} \sum_p \hat{f}_p q_p^{1'}$ and for $1 \leq i \leq r-1$, $\lambda_i = \frac{1}{c_0} \sum_p \hat{f}_p \left( \tilde{c}_i q_p^{i'} - \tilde{c}_{i+1} q_p^{i+1'} \right)$,

$$\sum_{p\in\mathcal{P}} \hat{f}_p d_p - \sum_{i=0}^{r} \tau_i \triangle_i \geq -\sum_{i=0}^{r} \lambda_i \tau_i, \qquad (16)$$

and substituting (16) into Lemma 15 gives the result. $\square$

*Proof of Lemma 17.* The proof is by induction on $r$. For $r = 0$, the lemma is satisfied at equality. Assume $\sum_{i=0}^{r-1} \lambda_i y_i \leq y_{r-1} \max_{k \leq r-1} \sum_{i \geq k} \lambda_i$. If $\sum_{i=0}^{r-1} \lambda_i y_i \leq 0$, then $\sum_{i=0}^{r} \lambda_i y_i \leq \lambda_r y_r \leq y_r \max_{k \leq r} \sum_{i \geq k} \lambda_i$.

If $\sum_{i=0}^{r-1} \lambda_i y_i > 0$, then by the induction hypothesis,

$$0 < \sum_{i=0}^{r-1} \lambda_i y_i \leq y_{r-1} \max_{k \leq r-1} \sum_{i \geq k} \lambda_i$$

and hence $\max_{k \leq r-1} \sum_{i \geq k} \lambda_i > 0$. Thus

$$\sum_{i=0}^{r} \lambda_i y_i \leq y_{r-1} \max_{k \leq r-1} \sum_{i \geq k} \lambda_i + y_r \lambda_r$$

$$\leq y_r (\max_{k \leq r-1} \sum_{i \geq k} \lambda_i + \lambda_r)$$

$$\leq y_r \max_{k \leq r} \sum_{i \geq k} \lambda_i$$

proving the lemma. □

*Proof of Lemma 18.* Since we now concentrate on a single phase $k$, we simplify notation and use $x'_e$, $l'_v$ and $q'_e$ to denote the rate flow, rate of change vertex labels and rate of change of queue in phase $k$. $E_1$ denotes the set of edges with strictly positive queues at the start of phase $k$. By definition for an edge $e = (v, w)$, $q'_e = l'_w - l'_v$ if $e$ is in the shortest path network and $q'_e = 0$ otherwise. Thus $c_e q'_e = x'_e(1 - \frac{l'_v}{l'_w})$ since if $e$ is not in the shortest-path network, $q'_e = x'_e = 0$. If $e$ is in the shortest-path network and $q'_e = 0$, $l'_v = l'_w$, and if $q'_e \neq 0$, $e \in E_1$ and hence $x'_e = c_e l'_w$. Thus $\sum_e c_e q'_e = \sum_{e=(v,w)} x'_e \left(1 - \frac{l'_v}{l'_w}\right)$. Let $\{x'_p\}_{p \in \mathcal{P}}$ be a path decomposition of $x'$. Then

$$\sum_e c_e q'_e = \sum_{e=(v,w)} \left(1 - \frac{l'_v}{l'_w}\right) \sum_{p \in \mathcal{P}: e \in p} x'_p$$
$$= \sum_{p \in \mathcal{P}} x'_p \sum_{e=(v,w) \in p} \left(1 - \frac{l'_v}{l'_w}\right). \qquad (17)$$

We bound $\sum_{e=(v,w) \in p} \left(1 - \frac{l'_v}{l'_w}\right)$ for any $s$-$t$ path $p$. Let $p = (s = v_0, v_1, v_2, \ldots, v_l = t)$, and $y = \sum_{i=0}^{l-1} \left(1 - \frac{l'_{v_i}}{l'_{v_{i+1}}}\right)$. Then

$$\frac{\partial y}{\partial l'_{v_i}} = \frac{l'_{v_{i-1}}}{(l'_{v_i})^2} - \frac{1}{l'_{v_{i+1}}}$$

for $i \neq 0, l$ and hence $y$ is maximized when $l'_{v_i} = \sqrt{l'_{v_{i-1}} l'_{v_{i+1}}}$. Note that by Lemma 2, $l'_v(\theta) \geq 0$ for any $v \in V, \theta \in \mathbb{R}_+$. We know that $l'_s = 1$, and by Lemma 11, $l'_t = c_0/\tilde{c}_k$. Substituting these values in the expression for $y$ yields

$$y \leq l - l \left(\frac{\tilde{c}_k}{c_0}\right)^{1/l}. \qquad (18)$$

Let $z = l - l \left(\frac{\tilde{c}_k}{c_0}\right)^{1/l}$ and $a = \frac{\tilde{c}_k}{c_0}$. Differentiating $z$ w.r.t. $l$,

$$\frac{\partial z}{\partial l} = 1 - a^{1/l} + \frac{a^{1/l}}{l} \ln a.$$

Hence, $z$ is maximized when $l - la^{1/l} = a^{1/l} \ln \frac{1}{a}$. Then $z = l - la^{1/l} = a^{1/l} \ln \frac{1}{a} \leq \ln \frac{1}{a}$, since $a \leq 1$. Substituting this in (18), $y \leq z \leq \ln \frac{c_0}{\tilde{c}_k}$. Hence on any path $p$, $\sum_{e=(v,w) \in p} \left(1 - \frac{l'_v}{l'_w}\right) \leq \ln \frac{c_0}{\tilde{c}_k}$. Substituting this in (17) yields

$$\sum_e c_e q'_e \leq \sum_{p \in \mathcal{P}} x'_p \ln \frac{c_0}{\tilde{c}_k} = c_0 \ln \frac{c_0}{\tilde{c}_k}.$$

since the rate flow $x'$ has value $c_0$. □